\documentclass[aps,prl,twocolumn,showpcs,showpacs,amsmath,amssymb]{revtex4}  
\usepackage{graphicx}  
\usepackage{dcolumn}   
\usepackage{bm}        

\begin{document}

\title{Ising models on the Regularized Apollonian Network}

\author{M. Serva$^{1,2}$, U. L. Fulco$^1$ and E. L. Albuquerque$^1$}

\affiliation{$^1$Departamento de Biof\'isica e Farmacologia, Universidade Federal 
do Rio Grande do Norte, 59072-970 Natal-RN, Brazil}
\affiliation{$^2$Dipartimento di Ingegneria e Scienze dell'Informazione e Matematica, 
Universit\`a dell'Aquila, 67010 L'Aquila, Italy}

\date{\today}

\begin{abstract}

We investigate the critical properties of Ising models on a Regularized Apollonian Network 
(RAN), here defined as a kind of Apollonian Network (AN) in which the connectivity asymmetry
associated to its corners is removed.
Different choices for the coupling constants between nearest neighbors are considered,
and two different order parameters are used to detect the critical behaviour.
While ordinary ferromagnetic and anti-ferromagnetic models on RAN do not undergo a phase 
transition, some anti-ferrimagnetic models show an interesting infinite order transition.
All results are obtained by an exact analytical approach based on iterative partial 
tracing of the Boltzmann factor as intermediate steps for the calculation of the 
partition function and the order parameters. 
\end{abstract}

\maketitle


Many real world networks exhibit complex topological properties as the 
small word effect,
related to a very short minimal path between nodes, and
the scale-free property, related to the power-law nature of the connectivity
distribution. 
These properties have important implications in the real phenomena
as virus spreading in computers, sharing of technological
information and diffusion of epidemic diseases, to name just a few.

In this context, the Apollonian network \cite{AHAS2005}
is a particularly useful theoretical tool,
since it is scale-free, displays
small-world effect, can be embedded in a Euclidean lattice and shows space-filling
as well as matching graph properties.
Therefore, in spite of its deterministic nature, it shares the most relevant 
characteristics of real world networks.

Phase transitions has been detected for a number of different physical models 
on Apollonian Networks.
For example, the ideal gas undergoes to Bose-Einstein condensation
\cite{CAS2008,0MLAA2009,0MLAA2010,0MCL2010,OSMLS2013}
and epidemics exhibits a transition between an absorbing state and an active state
\cite{SCSMFA2013,SCCMSFA2013}.
In particular, in \cite{OSMLS2013} it has been adopted an analytical strategy 
which has some similarities with that in this paper. 

In this work we focus on the infinite order transition exhibited by some
Ising models on the Apollonian network. Previous studies 
of similar Ising models \cite{AH2005, AAH2009} and Potts model
\cite{AAH2010} have not detected critical properties,
due to the fact that infinite order transitions are elusive.
On the contrary, a second order 
phase transition, as a function of the noise parameter, has been detected 
for a majority vote model \cite{LMA2012}.

Ising models on different hierarchical fractal have been studied \cite{GF1982} 
and in the case of diamond fractal they have been exactly solved
by exact renormalization \cite{DSI1983,BZ1989}.
These models, differently to present model,
show a second order phase transition.
Indeed, the standard Ising behavior is second order transition 
in plane models \cite{S2010,S2011},
nevertheless, it may be very intricate,
with many phases \cite{PS1997}, when the interactions are made more complicated.

We start by regularizing the standard Apollonian Network in order to remove
the connectivity asymmetry associated to its corners which
consistently simplify the analytical computation of the thermodynamics 
of the Ising models. 
 
The Regularized Apollonian Network ($RAN$) is defined starting form 
a $g=0$ generation network with 4 nodes all connected,
forming a tetrahedral structure with 6 bonds.
Each of the four triples of nodes individuates a different triangle.
At generation $g=1$ a new node is added inside each of
the four triangles and it is connected with
the surrounding three nodes,
 creating 12 new triangles. 
Then the procedure is iterated at any successive generation
inserting new nodes in the last created triangles,
and connecting each of them with the three surrounding nodes.

In $RAN$ the connectivity of any of the
already existing nodes (so-called old nodes) is doubled when generation 
is updated, while the connectivity of the
newly created nodes (the new nodes) always equals 3, leading to
 the following relevant property: 
the connectivity at generation $g$ of a node only depends on its age. 
More explicitly, its connectivity  is $3 \times 2^{g-g'}$ where $g'$ is the 
generation in which it was created. Besides, $RAN$ has the following properties:
(i) the total number of nodes is $N_{g}=(4 \times 3^{g}+4)/2$;
(ii) the number of new nodes created at generation $g \ge 1\,\,$ is 
$\,\, 4 \times 3^{g-1} \simeq (2/3) \, N_{g}$ 
(equality for large $g$); 
(iii) the average connectivity is $C_{g} = 2 U_{g}/N_{g} \simeq 6$ 
(for large $g$);
(iv) the total number of bonds is  $U_{g}=2 \times 3^{g+1}$; 
$v$) the number of new bonds created at generation $g \ge 1\,\,$ is 
$\,\, 4 \times 3^{g} = (2/3) \, U_{g}$.

The number of nodes having coordination $k$ is $m(k,g)$ which 
equals $\,4 \times 3^{g-g'-1}\,$ if $k=3 \times 2^{g'}$ with 
$g'=0,....,g-1$; equals 4 if  $k=3 \times 2^{g}$;
and equals $0$ otherwise. Accordingly,
the cumulative distribution
$P(k)=\sum_{k'\geq k} m(k,g)/N_{g}$ exhibits, for large values of
$g$, a power-law behavior i.e., $P(k)\propto {1}/{k^{\eta}}$,
with $\eta=\ln(3)/\ln(2) \simeq 1.585$.

Analogously to AN, RAN is scale-free and displays the small word effect.
Furthermore, since RAN network can be  decomposed in four AN networks cutting
a finite number of couplings, the thermodynamics on the two models is the same. 

Ising models are defined according to the following Hamiltonian:
\begin{equation}
H_g=-\sum_{i,j} J_{ij} \sigma_i \sigma_j
-\sum_{i} h_{i} \sigma_i - q \sum_{i,j, k}  \sigma_i \sigma_j \sigma_j,
\label{Hg}
\end{equation}
where the first sum goes on all $U_g/2$ connected pairs of nodes
of $RAN$ of generation $g$, the second sum goes on all $N_g$ nodes,
and the third sum only goes
on the $4\times 3^{g}$ triangles of the last generation $g$ 
(one of the nodes $i, j$ or $k$ must be lastly generated).
The constants $J_{ij}$ and $h_i$ may depend on the connectivities 
(on the age) of the involved nodes.
The couplings $J_{ij}$ can be both positive (ferromagnetic) or negative 
(anti-ferromagnetic).
The constant $q$ is introduced only for technical reasons, 
the relevant physics corresponding to $q=0$.

The partition function is
\begin{equation}
Z_g =\sum_{\#} \exp(-\beta H_g),
\label{Zg}
\end{equation}
where the sum goes on all $2^{N_g}$ configurations and $\beta$
is the inverse temperature, i.e., $\beta=1/T$ (we consider an unitary 
Boltzmann's constant $k_B$).
Then, the thermodynamical variables can be obtained from 
\begin{equation}
\Phi= 
\lim_{g \to \infty} (1/N_g) \log(Z_g).
\label{Phi}
\end{equation}

Our strategy consists in performing a partial sum in (\ref{Zg})
with respect to the $4\times 3^{g-1}$ spin variables
over nodes created at the last generation $g$. 
This sum creates new effective interaction between all remaining spins
and new magnetizations, and a new value for the parameter $q$.
In other words, we exactly map
the $g$ generation model in the same $g-1$ model with new parameters.
This technique works for any possible choice of the parameters $J_{ij}$,
$h_i$ and $q$, but we will consider here only some simple cases.

We stress that our approach is different when compared to the transfer matrix 
technique \cite{AH2005, AAH2009} and it gives exact expressions
for the thermodynamical variables in the $g \to \infty$ limit. 
While it confirms the absence of transition
in the ordinary ferromagnetic and anti-ferromagnetic models, 
it  detects an infinite order phase transition in 
a simple anti-ferrimagnetic model occurring at a finite temperature, 
in contrast with what was found
in \cite{AH2005, AAH2009}, where no critical behavior 
at a finite temperature was identified for this kind of models.

In order to illustrate our strategy we start with the simplest case
in which all interactions $J_{ij}$ are equal ($J_{ij}=J$), $q=0$
and all $h_i=0$.
Without loss of generality, one can chose $J=1$ 
(ferromagnetism) or $J=-1$ (anti-ferromagnetism).

Since a new node is only linked to three older surrounding nodes
(it is created inside a triangle),
the summation on the spin over a new node creates an extra-interaction among 
the three surrounding spins on the older nodes.
Therefore,  the partial sum on spins over new nodes in (\ref{Zg}) yields,
after some lengthy but straightforward calculations,  the following equality
\begin{equation}
 \log [Z_g(\beta, J)] =  \log [Z_{g-1}(\beta, J_1)]
+ 4 \times 3^{g-1} A(\beta),
\label{Zgf}
\end{equation}
where 
\begin{equation}
J_1= J + (1/2\beta)\log[2\cosh(2\beta)-1].
\label{J1} 
\end{equation}
Also,
\begin{equation}
A(\beta) = (1/4)\log[2\cosh(2\beta)-1] + \log[2\cosh(\beta)].
\label{A}
\end{equation} 
In (4), $Z_{g-1}(\beta, J_1)$  is the partition function of the same model
at generation ($g-1$) with a different value $J_1$ ($J_1$ has the same sign of $J$).
Note that in RAN (contrary to AN) equality (\ref{Zgf}) is exact. 

Performing the thermodynamical limit $g \to \infty$,
one obtains from (\ref{Zgf})
\begin{equation}
\Phi(\beta, J) =  (1/3)\Phi(\beta, J_1)+ (2/3) A(\beta),
\label{Phif}
\end{equation}
where we have used $N_{g-1}/N_g \to 1/3 $ and 
$4\times 3^{g-1}/N_g \to 2/3$.
We have thus re-expressed the thermodynamical function $\Phi(\beta, J)$
in terms of $\Phi(\beta, J_1)$, proving the absence of transition.
In fact, since $\Phi(\beta, J)$ only depends on the product $\beta J$,
and since the above equation can be iterated, a single
non-analytical point would imply an infinite number of 
non-analytical points.

Iteration of (\ref{Phif}) gives
\begin{equation}
\Phi(\beta ,J) =  
\frac{2}{3} \sum_{k=0}^\infty \frac{1}{3^{k}} A(\beta J_k)
\label{Phifs}
\end{equation}
where 
\begin{equation}
J_k=J_{k-1} + (1/2 \beta)\log[2\cosh(2\beta J_{k-1})-1]
\label{bk}
\end{equation}
with $J_0 = J$.
If $J_0 = J =1$ (ferromagnetism), the positive $J_k$ increase monotonically
and diverge for large $k$. On the contrary, if $J_0 = J =-1$ 
(anti-ferromagnetism) the negative $J_k$ converge to 0 for large $k$.
In both cases it is easy to verify that the sum (\ref{Phifs}) converges.
\begin{figure}[!ht]
  \includegraphics[width=2.5truein,height=2.5truein,angle=0]{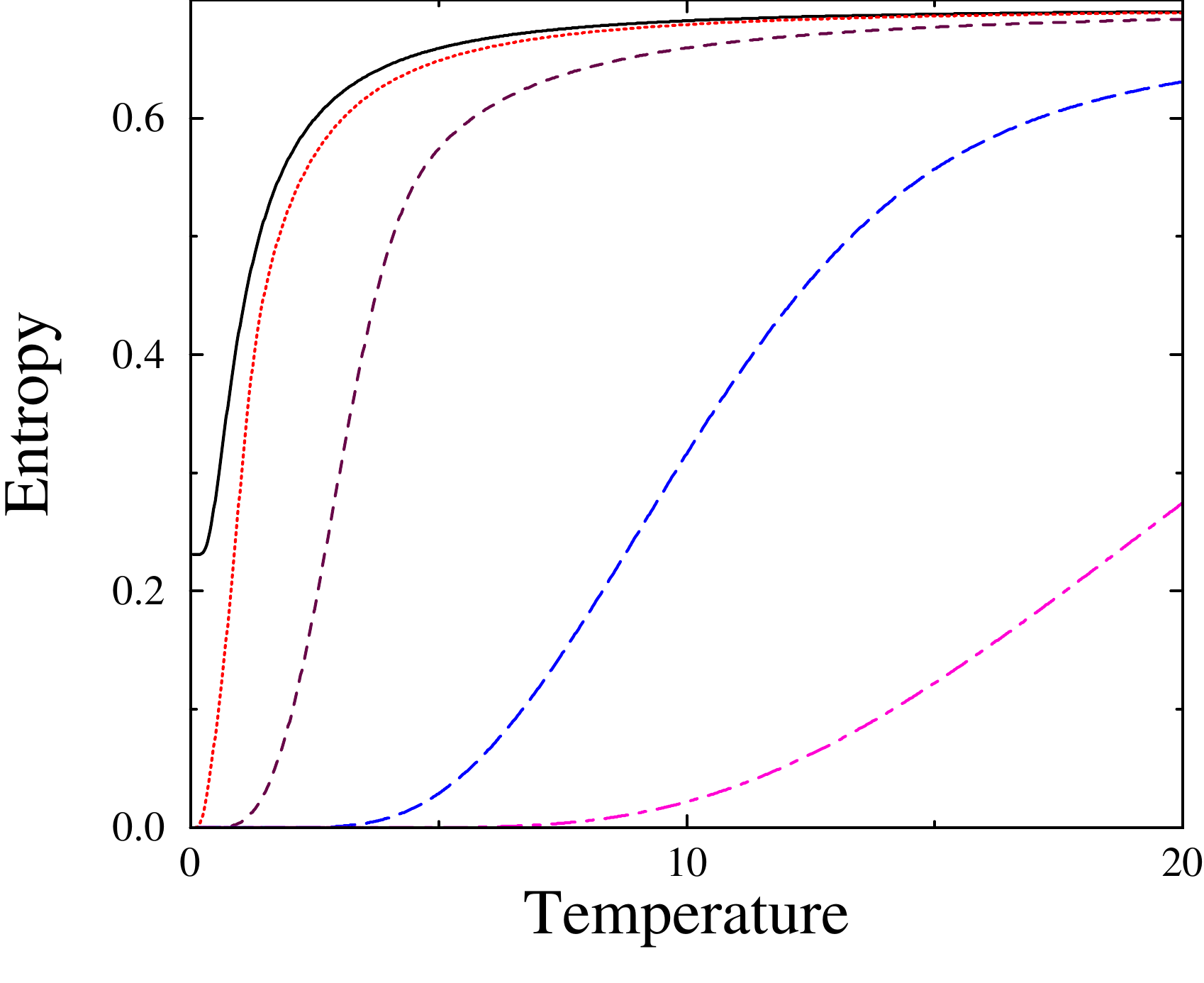}
  \caption{\small (color on line) Entropy versus the temperature $T$ (in units of the Boltzmann's constant $k_B$) for the following values of $u$ (from below): $u=10,5,2,1.2,1$. The case $u=1$ is the regular anti-ferromagnet
with non vanishing zero temperature entropy (entropy = $\log(2)/3$). 
Interestingly, whenever  $u>1$, the zero temperature entropy drops to 0. 
All curves behave regularly and do not show any sign of phase transition.}
  \label{fig1}
\end{figure}
\begin{figure}[!ht]
  \includegraphics[width=2.5truein,height=2.5truein,angle=0]{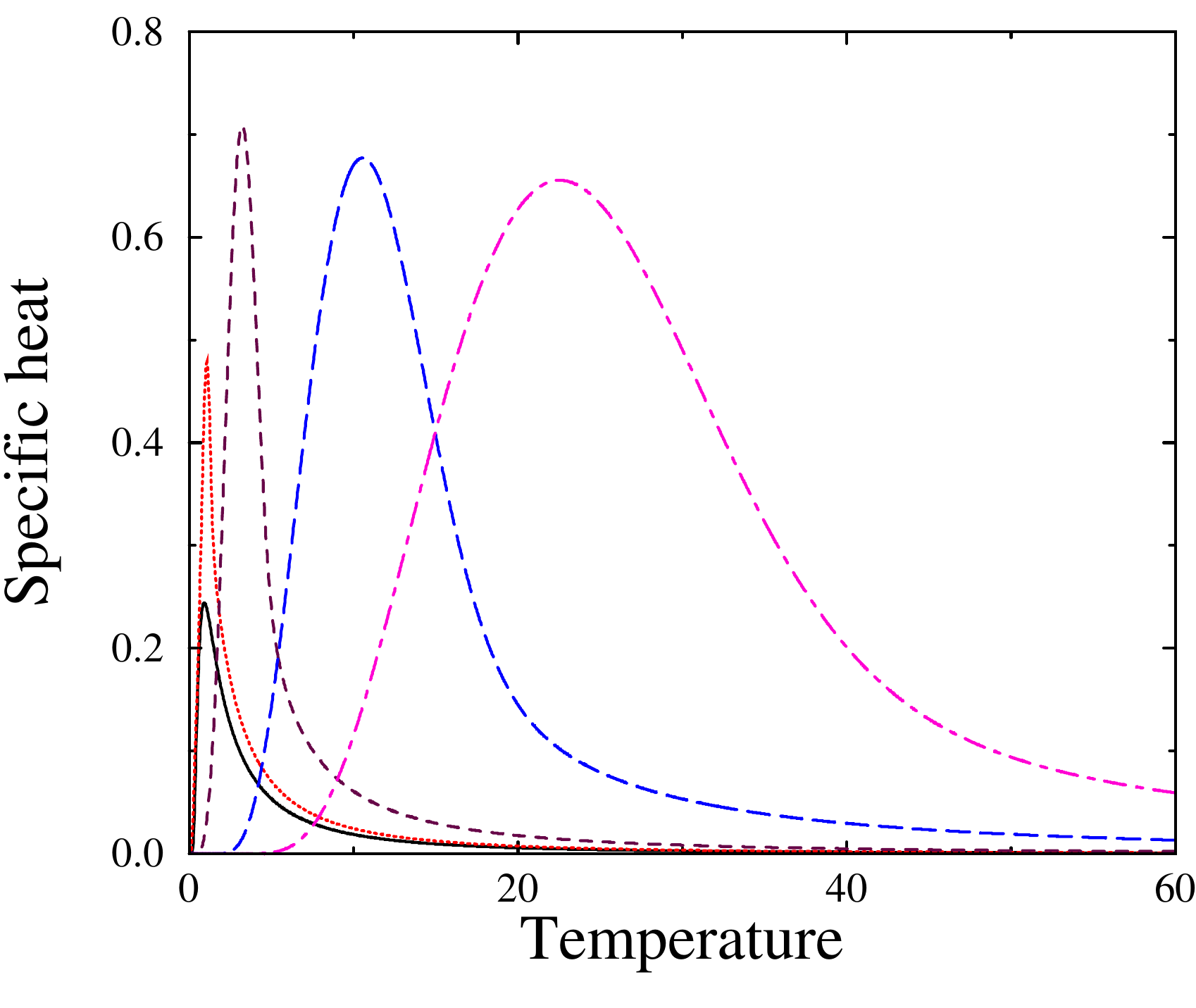}
  \caption{\small (color on line) Specific heat versus the temperature $T$ (in units of the Boltzmann's constant $k_B$) for the following values of $u$ (from the right): $u=10,5,2,1.2,1$. The curves behaves regularly and do not show any sign of phase transition.}
  \label{fig2}
\end{figure}

Since we have proven that the both constant coupling cases (the ferromagnet
and the anti-ferromagnet one) do not undergo a phase transition,
we now extend our scope to consider a different model.

Let us now come back to the Hamiltonian (\ref{Hg}) and let us assume that
the action of the external magnetic field $h_i$ is proportional to the connectivity of
the node, i.e., $h_i= h \, z_i$ where $z_i$ is the connectivity of node $i$. 
Accordingly, we can define the following spontaneous magnetization:
\begin{equation}
M= \lim_{g \to \infty} 
\frac{\sum_i z_i \langle \sigma_i \rangle }{\sum_i z_i} = 
\frac{1}{6 \beta} \left[ \frac{\partial\Phi}{\partial h} \right]_{h=q=0^+},
\label{M}
\end{equation}
where we have used $\sum_i z_i = 2 U_g \simeq 6 N_g$. 
The notation $\langle \cdot \rangle $ indicates average with
respect to the Gibbs measure, in such a way that $M$ satisfies $ 0 \le M \le 1$.
The rational for this choice  is that it turns out to 
be the simplest tool in order to show the existence of an infinite order 
phase transition in an anti-ferrimagnetic model.

For the same practical reasons, we compute the $accordance$ defined as 
\begin{equation}
L= \lim_{g \to \infty} 
\frac{\sum_{i,j, k}\langle \sigma_i \sigma_j \sigma_j \rangle }{4\times 3^g} 
= \frac{1}{2 \beta} \left[ \frac{\partial\Phi}{\partial q} \right]_{h =q=0^+},
\label{L}
\end{equation}
where the sum only goes on the $4\times 3^{g}$ triangles of last generation $g$. 
Note that the factor $1/2$ in the last term comes from the fact that 
in the limit $g \to \infty$ one has $4\times 3^{g}/N_g \to 2 $, 
accordingly $ |L| \le 1$.

Since we have proven that a constant value for the couplings $J_{ij}$ 
leads to absence of transition , we will assume now, on the contrary,
that they depend on the connectivity of the nodes $i$ and $j$ 
(in turn, the connectivity of a node  depends only on its age). 
The simplest age dependence for an anti-ferrimagnetic model
is $J_{i,j}=-u$ with $u>1$ if the connectivity
of at least one of the two nodes $i$ or $j$ is 3 and $J_{i,j}=-1$ otherwise. 
This is the same of assuming that $ J_{i,j}=-u$ for bonds involving 
nodes of last generation and $J_{i,j}=-1$ otherwise.
\begin{figure}[!ht]
  \includegraphics[width=2.5truein,height=2.5truein,angle=0]{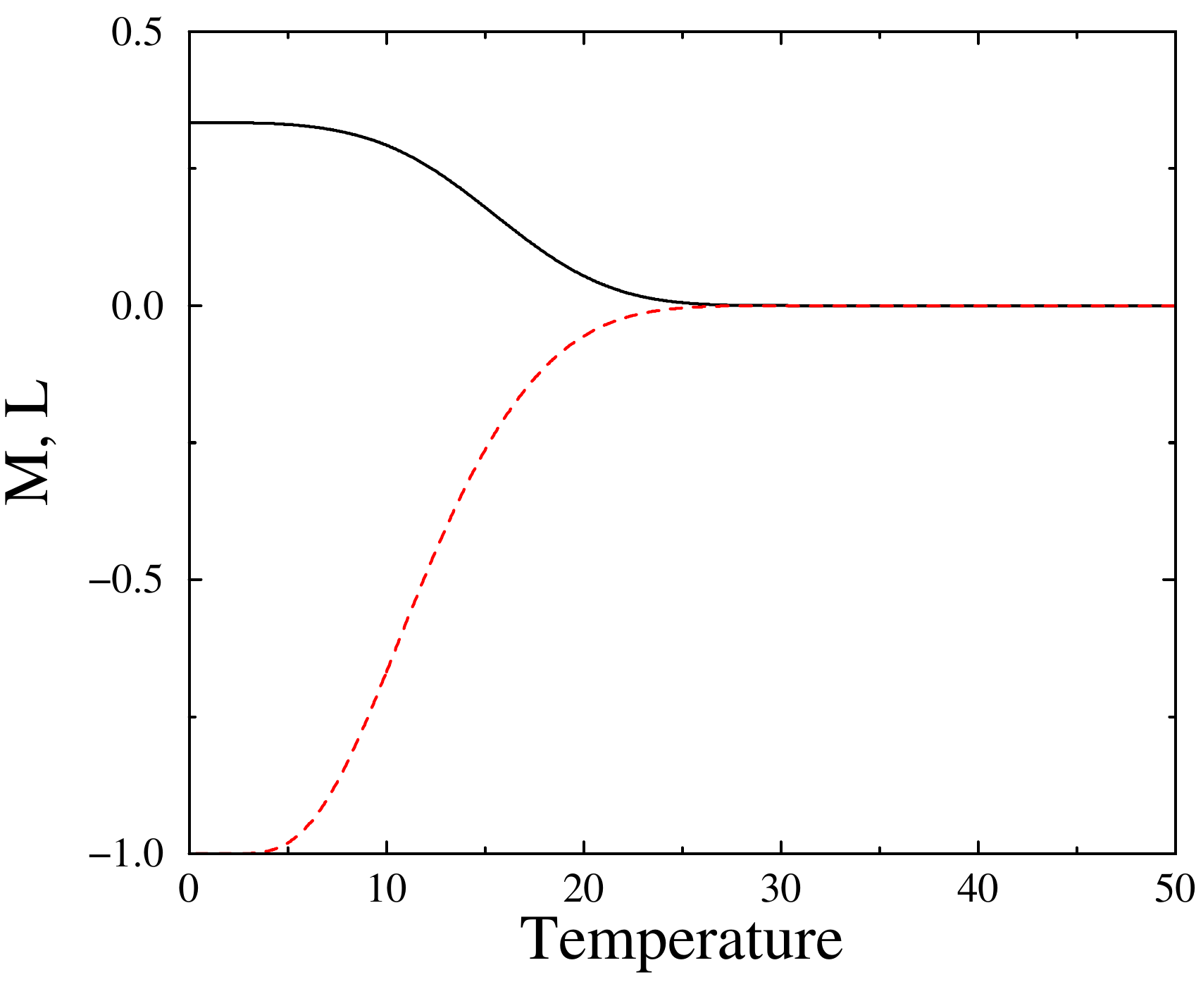}
  \vspace{0.5cm} \qquad
  \includegraphics[width=2.5truein,height=2.5truein,angle=0]{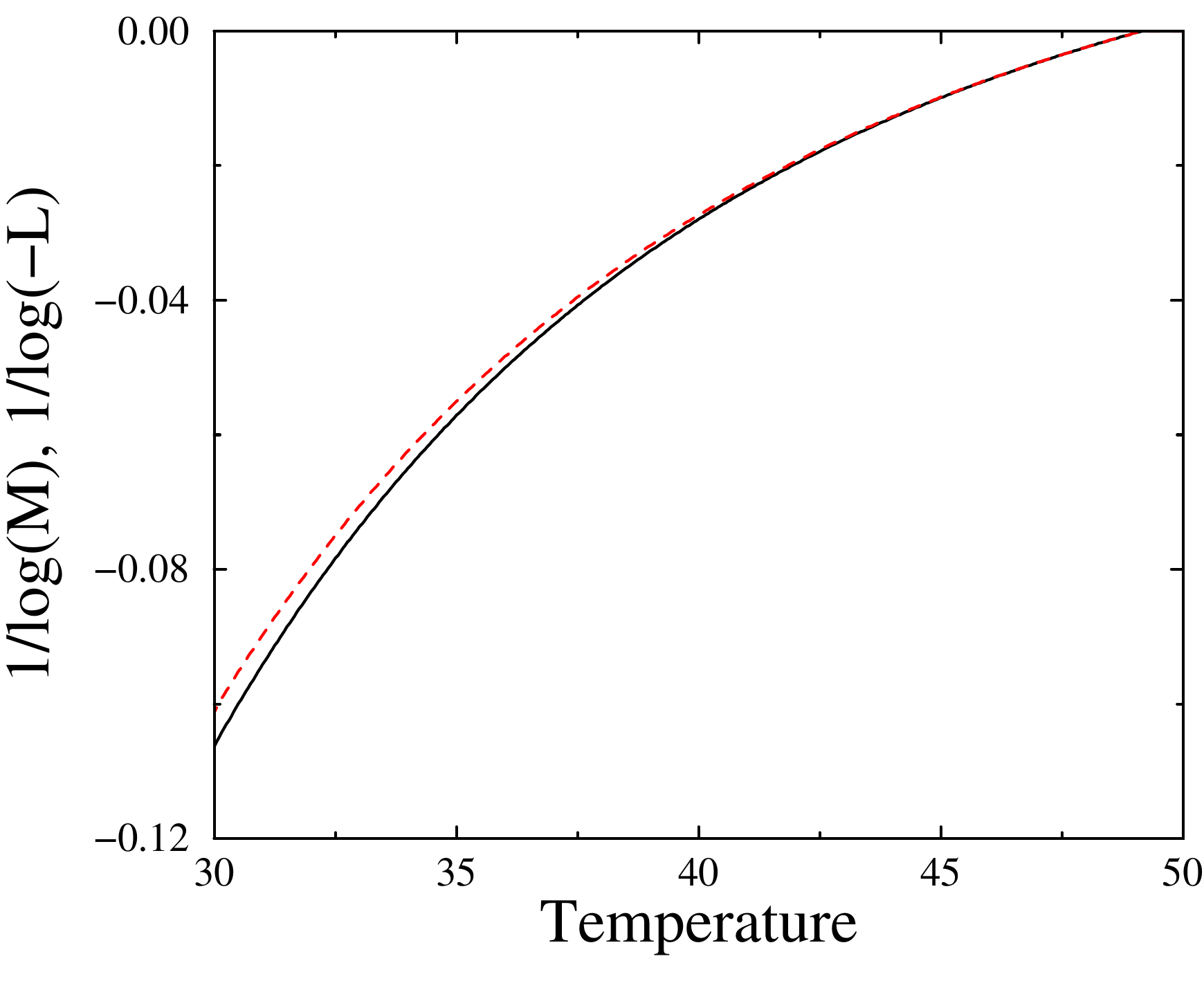}
  \caption{\small  (color on line) In the first panel are  depicted the spontaneous magnetization 
$M$ (full) and the $coordination$ $L$ (dashed) versus the temperature $T$ (in units of the Boltzmann's constant $k_B$) for the $u=5$ model.  
In the second panel are depicted $1/\log(M)$ (full) and $1/\log(-L)$ (dashed) versus the temperature $T$ (in units of the Boltzmann' constant $k_B$). 
In the first panel it seems that transition occurs at 
a temperature around 26, but this is a wrong perception.
The second panel shows, in fact, that the critical temperature 
is about  $T_c=49.16$ which is the correct transition temperature 
which we found analytically.
}
  \label{fig3}
\end{figure}
Then, following the same procedure, we take a partial sum with respect to the 
spins over last created nodes and we obtain the exact equality
\begin{equation}
\Phi(\beta, h, q, u) =  (1/3)\Phi(\beta, J_1, h_1, q_1)+ 
(2/3) A(\beta, h, q)
\label{bhq}
\end{equation}
We stress that, while the initial anti-ferromagnetic model 
(described by $\Phi(\beta, h, q, u)$) had two possible values for the couplings
($-u$ and $-1$), 
the effective model after partial sum (described by $\Phi(\beta, J_1, h_1, q_1)$)
has the single value $J_1$ for all of them.
The new parameters $J_1$, $h_1$, $q_1$ and $A$ can be again explicitly computed
in terms of $\beta, h, q, u$.
Since we are only interested in thermodynamical potentials, like the spontaneous magnetization 
$M$ and coordination $L$ in the absence of
external magnetic fields, what we really use are only
the values of $J_1$, $h_1$, $q_1$  for $q=h=0$ and their derivatives with respect to 
$h$ and $q$, in the limit $h \to 0$ and $q \to0$.

Assuming $h =0$ and $q = 0$ one has that  $h_1$ and $q_1$ are also equal to zero,
and we obtain again (\ref{Phifs}) with $A$ given by (\ref{A}).
The difference is that $J_0 = -u$,
\begin{equation}
J_1= -1 + (1/2\beta)\log[2\cosh(2\beta u)-1],
\label{beta1u} 
\end{equation}
while the remaining $J_k$ for $k \ge 2$ are again obtained by (\ref{bk}).
The resulting entropy and specific heat, as a function of the 
temperature $T$ (in units of the Boltzmann' constant $k_B$) 
are depicted in Fig.\ 1 and Fig.\ 2 for various values of $u$. 
Note that the case $u=1$ is the regular anti-ferromagnet 
with non vanishing zero temperature entropy. Interestingly, whenever 
$u>1$, as a consequence of the larger value of 
the coupling constants involving new nodes, frustration is removed and the zero 
temperature entropy drops to 0.  

Let us call $\beta_c= (1/2)\log[2\cosh(2u \beta_c)-1]$  the 
(non vanishing) value of $\beta$ for 
which $J_1$ vanishes, which only depends on $u$, 
then, there are three cases:
(i) $\beta=\beta_c$: in this case $J_1=0$ and one immediately obtains
$\Phi =  (1/3)\log(2)+   
(1/6)\log[2\cosh(2u\beta_c)-1] + 
(2/3) \log[2\cosh(u\beta_c)]$;
(ii) $\beta>\beta_c$: in this case $J_1>0$, i.e.,
the initial anti-ferrimagnetic model
is mapped into a ferromagnetic model.
The $J_k$ increases monotonically
and diverges for large $k$,
(iii) $\beta<\beta_c$: in this case $J_1<0$, i.e.,
the initial anti-ferrimagnetic model
is mapped into an anti-ferromagnetic model.
The $J_k$ converge monotonically to 0 for large $k$.
 
Is $\beta_c$  a critical point?  
The answer to this question is not easy since all 
thermodynamical functions seem to behave regularly at $\beta_c$
(see Figs.\ 1 and 2).
To address this question we have to compute the spontaneous 
magnetization $M$ and the coordination $L$.

Given $A(\beta, h, q)$ in (\ref{bhq}), it turns out that 
$(\partial A/\partial h)$
and $\partial A/\partial q$, as well as $(\partial J_1/\partial h)$
and $(\partial J_1/\partial q)$, calculated at $h=q=0$
vanish.
Then, given (\ref{M}) and (\ref{L}), we obtain from (\ref{bhq}):
$M=T_{11} \, M(J_1)+ T_{12}\, L(J_1)$ and
$L=T_{21} \,M(J_1)+ T_{22} \,L(J_1)$ where 
$M(J_1)$ and $L(J_1)$ are the magnetization and coordination
of the model with couplings $J_1$. Also, 

\begin{eqnarray}
T_{11}(\beta J_0)  =  \frac{1}{3}  \frac{\partial h_1}{\partial h} 
& = & \frac{2}{3}+ (1/4)[\tanh(3\beta J_0) + \tanh(\beta J_0)], \nonumber \\
T_{12}(\beta J_0)  =  \frac{1}{9} \frac{\partial q_1}{\partial h} 
& = & (1/12) [\tanh(3\beta J_0) -3 \tanh(\beta J_0 )], \nonumber \\
T_{21}(\beta J_0) = \frac{\partial h_1}{\partial q} 
& = & (1/4)[3\tanh(3\beta J_0) - \tanh(\beta J_0)], \nonumber \\
T_{22}(\beta J_0) =\frac{1}{3} \frac{\partial h_1}{\partial q} 
& = & T_{11}(\beta J_0) - \frac{2}{3},
\end{eqnarray}
where all derivative are calculated at $h=q=0$.

This relation can be iterated and one obtains
\begin{equation}
m = \prod_{k=0}^\infty T(\beta J_k) m(J_{\infty}),
\label{m}
\end{equation}
where $m$ is the vector whose two elements are $M$ and $L$;
$T(\beta J_k)$ are the $2 \times 2$ matrices with elements $T_{ij}(\beta J_k)$
and $m(J_{\infty})$ corresponds to spontaneous magnetization
and coordination of the model with couplings $J_{\infty}$.

If $\beta <\beta_c$, the sequence of $J_{k}$ converges to 0,
and $\prod_{k=0}^\infty T(\beta J_k)$ is a vanishing matrix;
therefore, $m=0$ independently on $m(J_{\infty})$.
If $\beta >\beta_c$, on the contrary, one find that
$J_{\infty} =\infty$, with both component of 
$m(\beta J_{\infty})$ equal to 1 (ferromagnet with infinitely large couplings).
Furthermore, in this case, $\prod_{k=0}^\infty T(\beta J_k)$ does not vanish,
and therefore  $m\neq 0$.
We have thus proven the existence of a transition at $\beta_c$.

The magnetization $M$ and the $coordination$ $L$ for the $u=5$ model
are depicted in the first panel of Fig 3.
Apparently, the transition occurs at a temperature around 26, which is far from the
transition temperature $T_c \simeq 49.16$ obtained by the non vanishing solution of 
$\beta_c= (1/2)\log[2\cosh(2u\beta_c)-1]$ with $u = 5$ and
$T_c=1/\beta_c$. 
This is due to the fact that both $M$ and $L$ vanishes extremely slowly.
In order to make this evident,  we have depicted $1/\log(M)$ and $1/\log(-L)$
in the second panel of Fig.\ 3,
which clearly show agreement with the expected critical temperature $T_c \simeq 49.16$.
The behaviour of both magnetization and $coordination$, close to the critical point, 
is compatible with the function $\exp[-C/(T_c-T)]$, which implies an infinite order 
transition since $M$ and $L$ and all their derivatives are continuous at $T_c$. 

Note that a negative coordination implies that spins are preferentially organized, 
in order that in any triangle one of them points in the opposite direction of the other two.
This organization becomes exact at $T=0$ where the coordination is -1.
Observe also that, contrary to the anti-ferromagnetic model on regular lattices, the
magnetization is positive below the critical temperature with a value compatible with
1/3 at $T=0$.

Qualitatively identical results come out whenever $u >1$,
leading to: (i) $\beta_c= (1/2)\log[2\cosh(2u\beta_c)-1]$
individuates the transition temperature; (ii) the transition is of infinite order.
In the limit  $u \to 1$ one easily  verify that $T_c =0$,
which confirms the absence of transition in the ordinary anti-ferromagnetic model.

Our minimal choice, $J_{i,j}=-u$ for newly created bonds
and $J_{i,j}=-1$ otherwise, 
is the simplest but it is not the only one which leads to
paramagnetic/ferromagnetic infinite order phase transition.
Indeed, there are many possible hierarchical choices
for the $J_{i,j}$ which lead to the same qualitative behavior.

The authors would like to thank the financial support from the Brazilian 
Research Agencies CAPES (Rede NanoBioTec and PNPD), CNPq 
[INCT-Nano(Bio) Simes, Casadinho] and FAPERN/CNPq (PRONEX). 
M.S. was partially supported by PRIN 2009 protocollo n. 2009TA2595.02.

\end{document}